\documentclass[12pt]{article}

\usepackage{url}  % Formatting web addresses
\usepackage{ifthen}  % Conditional
\usepackage{multicol}   %Columns
\usepackage[utf8]{inputenc} %unicode support
\usepackage{amsmath}
\usepackage{amssymb}
\usepackage{epsfig}
\usepackage{epstopdf}
\usepackage{graphicx}
\usepackage[margin=0.1pt,font=footnotesize,labelfont=bf]{caption}
\usepackage{setspace}
\usepackage{colortbl}
\usepackage[wide]{sidecap}
\usepackage[square,sort,comma,numbers,sort&compress]{natbib}
\usepackage{fullpage}
\usepackage{comment}
\usepackage{textcomp}
\usepackage{amsmath}
\usepackage[linesnumbered,lined,boxed,commentsnumbered]{algorithm2e}

\usepackage{algorithm2e}
\usepackage{algpseudocode}
\usepackage{lineno}
\usepackage{supertabular}
\usepackage{multirow}
\usepackage{booktabs}
\usepackage{longtable}

\usepackage{subcaption}

\usepackage{mdframed}
\definecolor{lgray}{rgb}{0.92,0.92,0.92}
\definecolor{lsalmon}{rgb}{1.0,0.63,0.48}

\DeclareUnicodeCharacter{00A0}{~}

\date{}

\begin{document}
\title{MarketGPT: Developing a Pre-trained transformer (GPT) for Modeling Financial Time Series}
\author{Aaron Wheeler and Jeffrey D. Varner\\ R.F. Smith School of Chemical and Biomolecular Engineering\\ Cornell University, Ithaca NY 14853\\}
\maketitle

\begin{abstract}
In this work, we present a generative pre-trained transformer (GPT) designed for modeling financial time series. The GPT functions as an order generation engine within a discrete event simulator, enabling realistic replication of limit order book dynamics. Our model leverages recent advancements in large language models to produce long sequences of order messages in a steaming manner. Our results demonstrate that the model successfully reproduces key features of order flow data, even when the initial order flow prompt is no longer present within the model's context window. Moreover, evaluations reveal that the model captures several statistical properties, or 'stylized facts', characteristic of real financial markets and broader macro-scale data distributions. Collectively, this work marks a significant step toward creating high-fidelity, interactive market simulations.
\end{abstract}

\section*{Introduction}
The recent success of Large language models (LLMs) has led to a boon in generative AI research across industries and disciplines. 
Financial time series modeling is a particularly promising research direction due to the vast amount of available financial market data. 
Conventional methods for modeling financial time series data (e.g., price and volume data) typically adopt a top-down approach and fit models to generate time series (e.g., price trajectories) directly \cite{Cox1979, Engle2001, Harvey1990, Stock2001, Black1973}. 
More recent approaches have utilized machine learning techniques, typically generative adversarial networks (GANs), to generate time series data directly \cite{Goodfellow2014, Eckerli2021}. 
However, these approaches struggle to reproduce all statistical properties of markets and typically abstract away valuable market microstructure features such as the limit order book (LOB). 
Several agent-based models have included these microstructure features along with agent behaviors of varying complexity, and they have demonstrated success reproducing certain statistical properties of financial markets \cite{Byrd2020, Lussange2022, Jericevich2021, Wheeler2023}. 
However, these models are difficult to calibrate and have struggled to reproduce complex statistical properties related to order flow. 
Only recently has there been work demonstrating the use of autoregressive generative models for bottom-up financial market microstructure simulation \cite{Jain2024, Hultin2023, Nagy2023}.

In this study, we developed a world agent to generate order flow within a interactive discrete event simulator (DES). 
Building on recent work, we model order flow using the same techniques used in language modeling: we tokenize individual components of order messages and train a world agent model to predict the next token (message component) in the sequence \cite{Webster1992}. 
We utilized a modern Transformer architecture, the backbone of recent LLMs, to model the world agent \cite{Vaswani2017}. 
We demonstrate that the model is capable of producing realistic order flow at nanosecond precision by evaluating the generated messages against well known statistics of market phenomena (i.e., stylized facts) \cite{Cont2001, Chakraborti2011, Gould2013, Vyetrenko2020}. 
By serving as a message generation engine inside of a DES, our model enables future work in studying financial markets and applications such as stress testing trading strategies and evaluating proposed market legislation ex-ante.

\clearpage

\section*{Materials and Methods}

\subsection*{Data}
We utilized historical Nasdaq TotalView-ITCH 5.0 message data to train our model. 
Nasdaq ITCH message data provides detailed information on every quote and order at each price level for securities traded on Nasdaq. 
We had access to 8 days of data, which we divided into 6 days for training, 1 day for validation, and 1 day for testing. 
The specific days reserved for training, validation, and testing varied by ticker symbol, which allowed us to assess our model's performance across different days. 
To process the data, we reconstructed the messages from the raw feed following the ITCH 5.0 protocol \cite{martinobdl2021}. 
We excluded all pre-market (before 9:30 AM ET) and after-hours messages (after 4:00 PM ET) and included messages from all price levels of the order book. 
This comprehensive price level inclusion was different from previous studies \cite{Nagy2023, Frey2023, Coletta2023, Hautsch2011} and was done to account for messages deeper in the book which may contain valuable information, such as cancellation dynamics and order book imbalances. 
These messages are especially significant during periods of significant price movement, like flash crashes, and are necessary for users of our simulation platform who wish to evaluate quote placements deeper in the book. 
Maintaining a full level-3 representation of orders across the book was also vital for evaluation purposes, such as estimating functional forms and computing distribution tails.

Our use of ITCH message data included six core elements: (i) timestamp at which the message was generated (in nanoseconds since midnight), (ii) message type, (iii) order ID, (iv) order direction (buy or sell), (v) order size, (vi) price. 
We utilized five message types: new limit order, execution of visible limit order (full or partial), execution of visible limit order in whole or in part at a price different from the initial display price (e.g., pegged order), cancellation of limit order (full or partial), and replacement of limit order. 
Some messages make multiple uses of the same core elements depending on the message type; for example, replace messages have an order ID field dedicated to both new and old order IDs, and cancel orders have an order size field dedicated to the canceled size and the size remaining after the deletion (may be total or partial). 
All message types, except new limit orders, contained referential order message information. Notably, we included the replace order message type, which is often absent in other pre-processed datasets and microstructure models \cite{Huang2011}. 
We excluded all other fields from the raw message data, such as the execution of hidden limit orders (since they do not affect the modeling of the LOB), market participant IDs, and event information (e.g., auctions and trading halts).

Before messages were tokenized, they were pre-processed to better suit deep learning tasks. 
This included transforming messages into more stationary representations as detailed in Nagy et al. \cite{Nagy2023}. 
Following this framework, we converted the price fields from dollar values to ticks from the previous mid-price $p_{t-1}$ and added inter-arrival times between messages as features. 
Rare events, such as price ticks quoted past 999 ticks away from the mid-price and order sizes above 9999, were truncated to those limits to maintain a finite vocabulary size, affecting only a small number of messages. 
Diverging from this framework, given that our dataset included the execute at different price and replacement message order types, we extended the framework to include the \emph{old ID} and \emph{old absolute price} fields. 
Further, the \emph{reference relative price}, $p^{ref}$, field had to be altered for these two order types. 
Rather than refer to the relative price of the matching limit order, the field contains the value of the price difference relative to the new mid-price at the time the message was sent. 
This allowed us to compute the old absolute price $p^{old}_{abs} = p_t + p^{ref}$ during inference and search the LOB price levels for the order to be replaced. 
Lastly, since we dealt with multiple different stocks in this work, we created a symbol mapping for each ticker and appended it to each message in the sequence. 

\subsection*{Tokenization}
\begin{figure}[t]
    % \centering
    \includegraphics[width=0.99\textwidth]{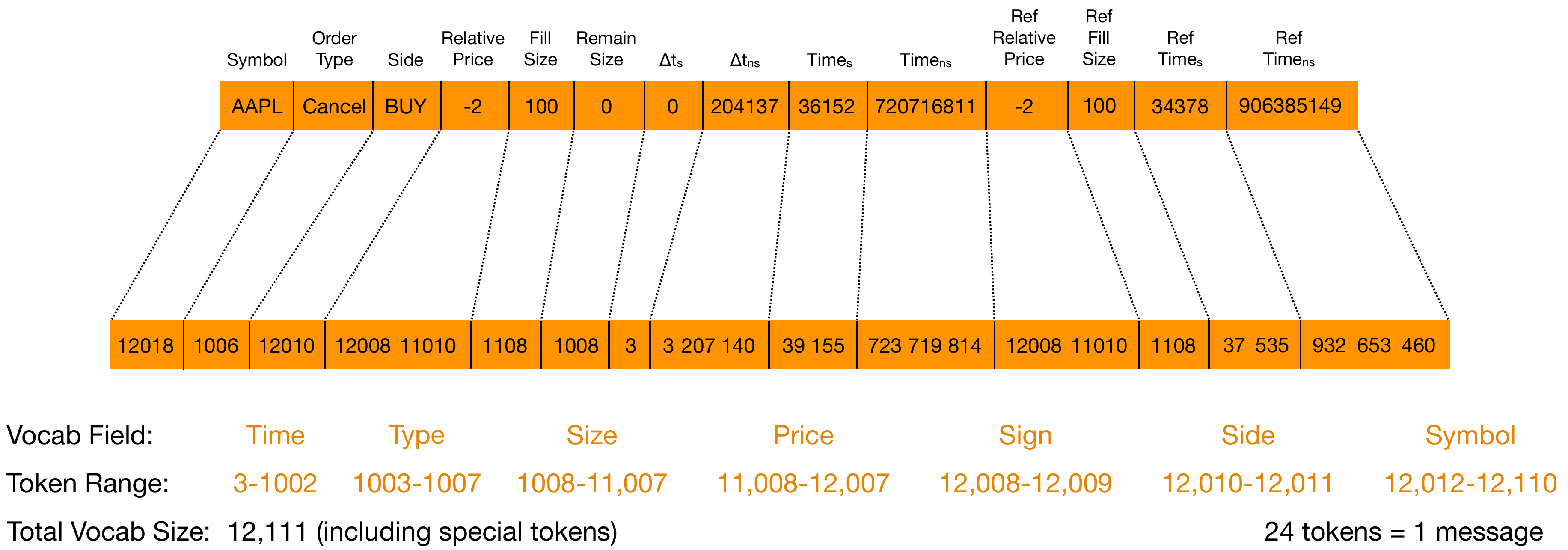}
    \caption{Tokenization scheme that translates pre-processed messages (top) into encoded messages (bottom). The vocabulary fields and corresponding valid token values are listed below the diagram. Pre-processed message fields that are not tokenized (e.g., order ID and absolute price) are excluded from the figure. Fields that are similar (e.g., $\Delta t$ and time) share vocabulary values. The first three tokens are reserved for special tokens:  masking tokens, $NaN$ tokens, and sink tokens. Reference message fields are $NaN$ for limit orders and correspond to the matching limit order for all other message types.}
    \label{fig:gpt_tokenizer}
\end{figure}

We built upon the tokenization scheme developed by Nagy et al. \cite{Nagy2023} to encode message features into real values which can be understood by our model. 
This set of values, or \emph{vocabulary} of the model, encodes the valid range of possible values each message field can be translated into. 
This was done to establish a finite and working set of tokens to predict from: the model learns to turn this vocab set into probability distribution over the next possible token. 
The tokenization process converts an 18-element length pre-processed message into a 24-element length tokenized message (Fig. \ref{fig:gpt_tokenizer}). 

% \begin{figure}[t]
%     % \centering
%     \includegraphics[width=0.99\textwidth]{Figures/marketGPT/MarketGPT_tokenization_fixed.pdf}
%     \caption{Tokenization scheme that translates pre-processed messages (top) into encoded messages (bottom). The vocabulary fields and corresponding valid token values are listed below the diagram. Pre-processed message fields that are not tokenized (e.g., order ID and absolute price) are excluded from the figure. Fields that are similar (e.g., $\Delta t$ and time) share vocabulary values. The first three tokens are reserved for special tokens:  masking tokens, $NaN$ tokens, and sink tokens. Reference message fields are $NaN$ for limit orders and correspond to the matching limit order for all other message types.}
%     \label{fig:gpt_tokenizer}
% \end{figure}

Notably, some fields (such as time and price) require multiple tokens to be encoded. 
The price fields require both a sign value (bid or ask side of book) and relative price value (distance from mid-price). 
The time field is split into seconds and nanoseconds components for both the $\Delta{t}$ field and message timestamp field. 
Although messages may contain fields with different underlying meaning ('size' can refer to fill size or remaining size), the same vocab value range is used for each variant of the same basic components. 
The vocab size for message encoding is 12,012 + $S$, where $S$ is the number of tokens reserved for ticker symbol IDs. In this study, we reserved room for $S=98$ tickers for total vocab size of 12,111 tokens. 
During inference, messages are dynamically encoded (for the model to process) and decoded (for the simulator to process).

\subsection*{Transformer based world agent}
The model architecture was inspired by recent advancements in Transformers, which have become the cornerstone of deep learning success in language modeling \cite{Zhao2023}. 
Recent improvements have made Transformers more computationally efficient and accessible for use with modest hardware. The model was developed using the PyTorch deep learning framework \cite{PyTorch}. 
Our model design included 768 embedding dimensions, 12 layers, and 12 heads, totaling approximately 100 million parameters. 
The input to the model consisted of message tokens $\textbf{x}$ with a maximum sequence length $\mathcal{L}$ (context size). 
Specifically, inputs $\textbf{x} = \left\{x_0,\cdots,x_{\mathcal{L} - 1}\right\}$ and labels $\textbf{y} = \left\{x_1,\cdots,x_{\mathcal{L}}\right\}$ were derived from flattened sequences of $n$ tokenized messages $\textbf{m} \in \mathcal{V}^{24n}$, where $\mathcal{V} \subset \mathbb{N}$ denotes the token vocabulary. 
Similar to the language modeling task, the training objective was to autoregressively predict the target tokens $x_i$ based on the preceding tokens $x_{<i}$ in the sequence. 
This is done efficiently by the decoder-only transformer architecture and unidirectional attention mask allowing parallel computation across the sequence length. 
The model $f_{\theta}: (\textbf{x}) \mapsto \hat{\textbf{y}}$ was trained to map tokens sequences to a vector of logits $\hat{\textbf{y}} \in \mathbb{R}^v$ where $v = \left|\mathcal{V}\right|$ denotes the size of the token vocabulary. 
By applying the softmax operation to convert logits into probabilities, the model defined the conditional distribution of the next token, given the input sequence $P(x_i| \textbf{x}_{< i} )$. 
The model parameters $\theta$ were optimized by minimizing the cross-entropy loss over batches of training data using gradient descent with the Adam optimizer \cite{Kingma2014}.  

% include training parameters somewhere (appendix)? (including bfloat16)

Our training process showed significant improvements in training and validation loss when the model was pretrained using tokens from multiple tickers in the dataset. 
Following pretraining, the model was fine-tuned for a specific ticker symbol using data exclusively for that symbol. 
This fine-tuned model was then selected and deployed for inference. 
The pretraining dataset comprised 91,124,274 messages (2,186,982,576 tokens) across 20 assets, and the model was trained with a maximum sequence length $\mathcal{L}$ of 10,368 tokens ($n=$432 messages). 
Gradient accumulation was used to simulate a larger batch size, and flash attention was employed to reduce training times \cite{Dao2022}. 
A random grid search over hyperparameters was conducted to identify optimal training parameters \cite{Bergstra2012}.

To enhance model performance during training and inference, several techniques were applied. 
Root Mean Square Layer Normalization (RMSNorm) replaced LayerNorm in the original transformer implementation, retaining the re-scaling invariance property without the overhead of re-centering \cite{zhang-sennrich-neurips19}. 
Relative positional embeddings (RoPE) were used instead of traditional absolute positional encoding, providing benefits such as flexibility in sequence length, decaying inter-token dependencies, and enhanced self-attention \cite{Su2021}. 
Although we implemented Grouped-Query Attention (GQA) to balance performance and computational efficiency, it was not used in the final model due to its poor impact, likely due to the small model size \cite{Ainslie2023}.  
To accelerate inference, a KV Cache was implemented to eliminate the need to recompute key and value tensors in self-attention computations \cite{Pope2022}. 

Given the autoregressive and long-range nature of generating order flow across an entire trading day, the number of messages used as input can quickly exceed practical context length limits. 
Handling inputs that extend beyond the pre-trained context length is a problem also faced by LLMs, as research has shown that models degrade past this point and lose "fluency" \cite{Press2022, Chen2023}. 
Transformers further suffer from attention scaling quadratically with sequence length, complicating long-range problem handling. 
A straightforward approach to mitigate these concerns is limiting the tokens fed to the model using a constant-sized sliding attention window \cite{Beltagy2020}. 
However, Xiao et al. \cite{Xiao2023} found that models lose fluency immediately after the first token is evicted from the attention window and proposed using "attention sinks" with a sliding window to address this. 
Attention sinks function by offering a dedicated token for offloading attention scores when computing the next token to be generated. 
The intuition behind this approach is that models learn to use the first few tokens in a sequence for offloading attention scores (since softmax must be summed to one regardless of how relevant the previous tokens are) and attention sinks serve as a mechanism for the model to do so once the initial context length is surpassed. 
We found that the use of a single dedicated sink token during training and inference time greatly improved the fidelity and stability of our model, allowing it to be used for sequences greatly surpassing the pre-trained sequence length. Our use of a single dedicated sink token, along with RoPE and a rolling KV cache, enabled effective processing of messages in a streaming fashion.

We experimented with various sampling techniques, including truncation, temperature scaling, and nucleus sampling \cite{fan-etal-2018-hierarchical, Holtzman2020}. 
We found that using a temperature above 1.0 (i.e., scaling logits before applying softmax) was necessary to avoid overly conservative message generation (e.g., underprediction of low probability events such as execution order types and large order sizes and relative prices). 
We found that explicitly truncating the logits distribution to the $k$-highest probability tokens (top-k sampling) was prone to sampling errors due to the optimal value of $k$ being different for different output distributions (e.g., predicting order type token vs. relative price token). 
Nucleus filtering avoids the problem of choosing the optimal $k$ value altogether, and instead restricts sampling to the set of tokens with cumulative probability $<$ $p$ (top-p sampling). 
We found that a temperature of 1.02 and a $p$ value of 0.98 balanced simulation fidelity and rare event generation. 
We also limited sampling to the target token distribution for each token, which was done to avoid any sampling errors across long-range simulation trials. 

\subsection*{Discrete event simulator}
\begin{figure}[ht]
    % \centering
    \includegraphics[width=0.99\textwidth]{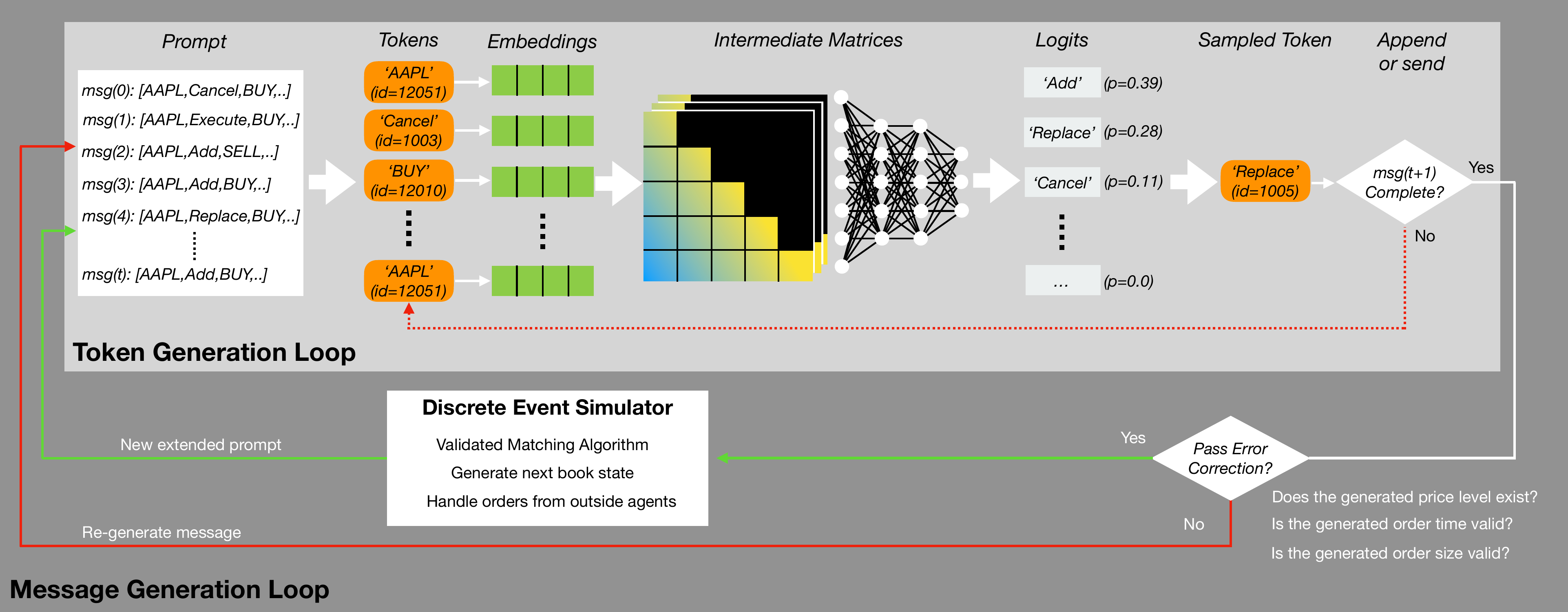}
    \caption{Schematic of the simulation platform. The token generation loop contains the prompt (previous messages up to context length) and the world agent---i.e., the transformer model. Upon sampling the encoded message length (24 tokens), the error correction procedure checks for and remedies hallucination conditions if possible. If the generated message is deemed valid by the error correction procedure, the discrete event simulator coordinates the message to an exchange agent, who then submits the order to the limit order book (LOB) and updates the LOB state. The message is then appended to the prompt and the cycle is repeated.}
    \label{fig:platform_schematic}
\end{figure}

To dynamically generate messages and construct LOB states, the DES would host at minimum two agents: a transformer based world agent to conditionally generate messages and an exchange agent to receive order messages and store and update the order books. 
The DES was based on the ABIDES simulator framework \cite{Byrd2020, Amrouni2021}. Each simulation trial was set to begin, at minimum, approximately 30 minutes after market open ($\sim$10:00am EST) for the test date in question. 
This was done to make things easier for the analysis of the model: the trading behavior that follows market open is typically much more active and volatile before settling into steady-state. 
Before the model generates any new messages, the order books hosted by the exchange agent were instantiated using all historical messages up to that point (including pre-market messages). 
The most recent context size length of messages were used as the initial "prompt" for the model. 
At inference time, the model would use this context to generate tokens in an autoregressive fashion until a new message was completed (i.e., when 24 tokens were generated). 
Similar to Nagy et al. \cite{Nagy2023}, newly generated messages that included a referential component would be subjected to an error correction procedure. 
The error correction procedure was implemented to handle "hallucinated" messages, which were messages that would refer to orders that did not exist (i.e., a cancel order message referring to a buy order quoted at \$99.00 when no such order existed in the book). 
First, the simulator would check whether or not the reference order price corresponded with an existing price level in the order book. 
If the price level did exist, then the simulator would check whether or not the generated order time or order size was valid---if neither condition was met then the first order in the priority queue of that price level was returned. 
If a message did not pass the error correction procedure, then the message was discarded and the simulator would re-run the timestep. 
This would occur for a relatively low number of messages ($\sim$7\%).
The vast majority of messages would pass error correction and be handled by the exchange agent, which would result in an updated LOB state. 
The message would then be appended to the context (unless the context length was exceeded, in that the case the first message would be evicted before appending the new message) and the process would repeat until the simulation was stopped (Fig. \ref{fig:platform_schematic}). All simulation trials were run on a single NVIDIA RTX 4090 GPU.

% \begin{figure}[ht]
%     % \centering
%     \includegraphics[width=0.99\textwidth]{Figures/marketGPT/MGPT_platform_schematic.pdf}
%     \caption{Schematic of the simulation platform. The token generation loop contains the prompt (previous messages up to context length) and the world agent---i.e., the transformer model. Upon sampling the encoded message length (24 tokens), the error correction procedure checks for and remedies hallucination conditions if possible. If the generated message is deemed valid by the error correction procedure, the discrete event simulator coordinates the message to an exchange agent, who then submits the order to the limit order book (LOB) and updates the LOB state. The message is then appended to the prompt and the cycle is repeated.}
%     \label{fig:platform_schematic}
% \end{figure} % ...state variables are updated (e.g., sim time, book) and other orders are handled before cycle repeats

\section*{Results and Discussion}
The model was evaluated by comparing generated messages to the target distribution (unseen test data---the real historical messages). Specifically, we evaluated outputs for the existence of several well-known statistical regularities at the message-level and the predictive potential of the resulting price and volume trajectories.
% .. the ability for the model to reproduce unconditional marginal distributions
% ... the ability for the model to reproduce conditional distributions (compare correlations between gen and real returns)
We used a model finetuned on AAPL data for evaluation. To evaluate this model, we chose the best combination of inference sampling parameters and generated sequences of approximately 20 simulated minutes of data. We compared these messages against the same number of real messages produced after the final context message. Due to hallucination, the generated sequences end up with different lengths---for evaluation purposes, we restrict series to a common sequence length of $T=135,813$ messages. Further, the pegged order message type was extremely rare in the training data and was not generated by our model so we excluded it from our analysis. We also decreased the model context length to $\mathcal{L}=2,688$ messages to reduce inference (wall-clock) times. We did not observe any substantial decrease in model performance from this context length reduction.

% effect of including pre-training, training curves? (appendix?)

% % order type percentages
% \begin{figure}[ht]
%     % \centering
%     \includegraphics[width=0.99\textwidth]{Figures/marketGPT/msg_type_percentage_12302019_AAPL.pdf}
%     \caption{Generated and empirical message type frequencies. Data across all simulation trials (N=10) were compared against the test distribution---error bars account for 95\% confidence intervals. The frequencies roughly match, with the only discrepancies occurring for add and replace message types.}
%     \label{fig:msg_percents}
% \end{figure}

\subsection*{Simulated order flow}
\begin{figure}[ht]
    \centering
    \begin{subfigure}{0.65\textwidth}
        \centering
        \includegraphics[width=\textwidth]{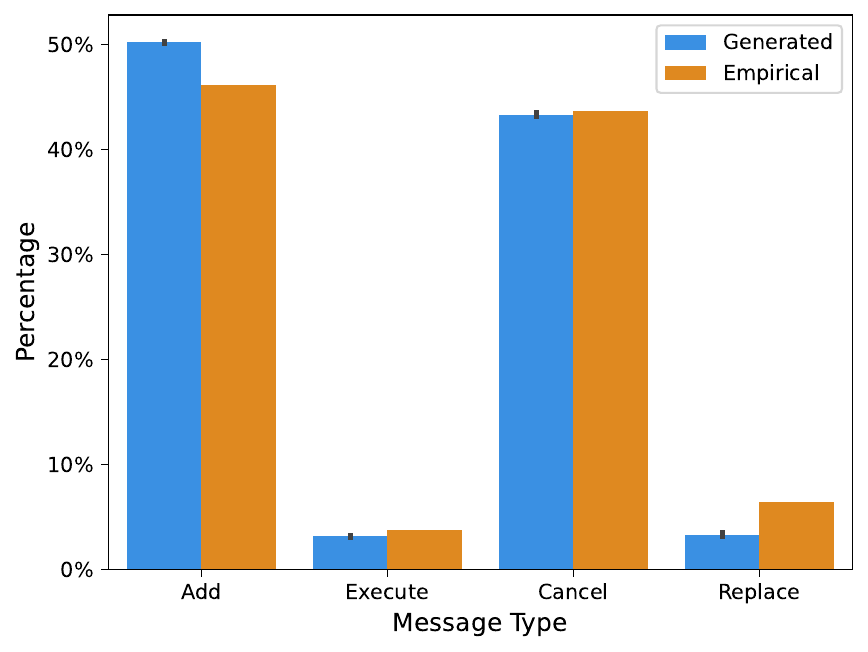}
        \caption{}
        \label{fig:msg_percents}
    \end{subfigure}
    \hfill
    \begin{subfigure}{0.65\textwidth}
        \centering
        \includegraphics[width=\textwidth]{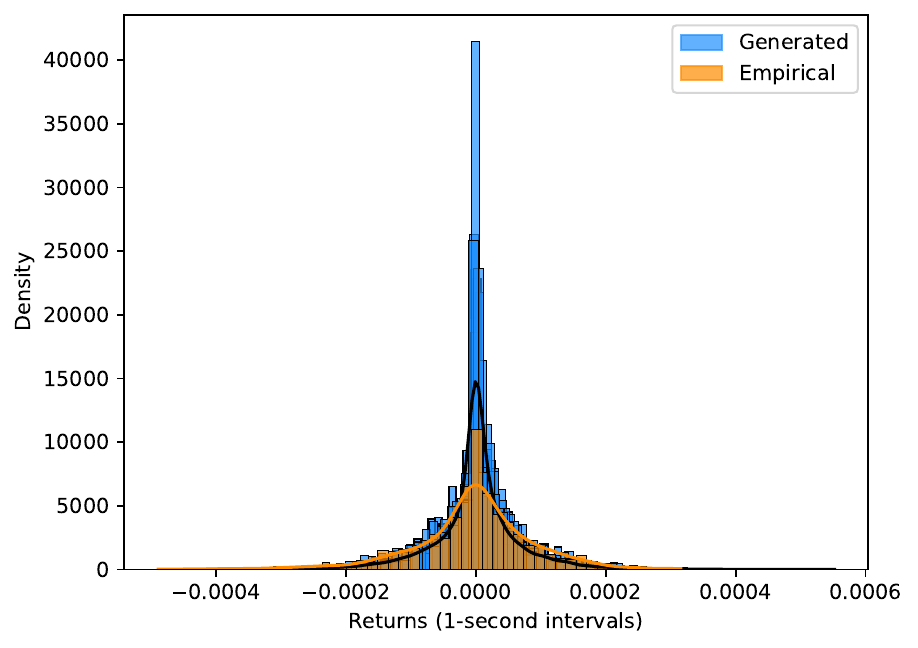}
        \caption{}
        \label{fig:heavy_tails}
    \end{subfigure}
    \caption{(a) Generated and empirical message type frequencies. Data across all simulation trials (N=10) were compared against the test distribution---error bars account for 95\% confidence intervals. The frequencies roughly match, with the only discrepancies occurring for add and replace message types. (b) Distribution of 1-second returns of both generated and empirical messages. Data across all simulation trials (N=10) were compared against the test distribution. The solids lines denote the kernel density estimate for both generated (black line) and empirical (orange line) returns. Both distributions exhibited the heavy-tailed property--however, the generated returns distribution was noticeably more extreme.}
    \label{fig:percents_tails}
\end{figure}

% order type frequency distribution
Order type frequencies produced by the model roughly matched empirical counterparts, with the only noticeable difference being that the model tended to slightly over-predict the number of limit orders and under-predict the number of replace orders (Fig. \ref{fig:msg_percents}). We attributed this discrepancy to the model's difficultly in predicting referential components of replace messages. 
Upon investigating the order placement errors caught by the simulator's error correction procedure, we uncovered that a disproportionate amount of these errors came from replace orders. We reasoned that since replace orders must also predict a valid absolute replace price (which cancel orders do not have to do), this left greater probability of error when replace orders were generated. Whenever this replace order placement error occurred, the message was discarded and it is likely that a limit order was placed next (since it is the most abundant message type in the training data)---so this may explain why limit orders are placed more frequently at roughly the same frequency that replace orders are placed less.

\begin{figure}[ht]
     \centering
     \begin{subfigure}{0.49\textwidth}
         \centering
         \includegraphics[width=\textwidth]{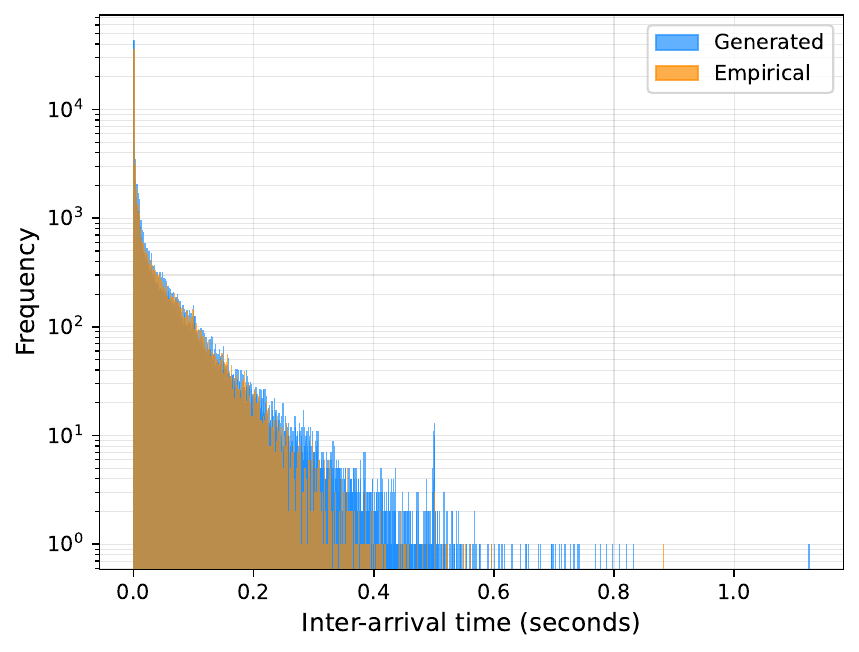}
         \caption{}
         \label{fig:add_inter}
     \end{subfigure}
     \hfill
     \begin{subfigure}{0.49\textwidth}
         \centering
         \includegraphics[width=\textwidth]{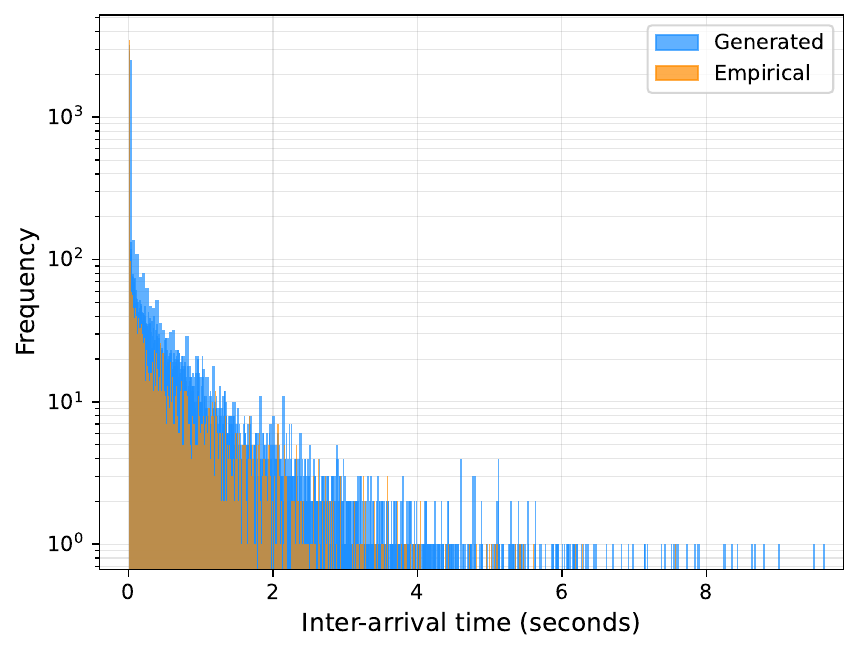}
         \caption{}
         \label{fig:exec_inter}
     \end{subfigure}
     \hfill
     \begin{subfigure}{0.49\textwidth}
         \centering
         \includegraphics[width=\textwidth]{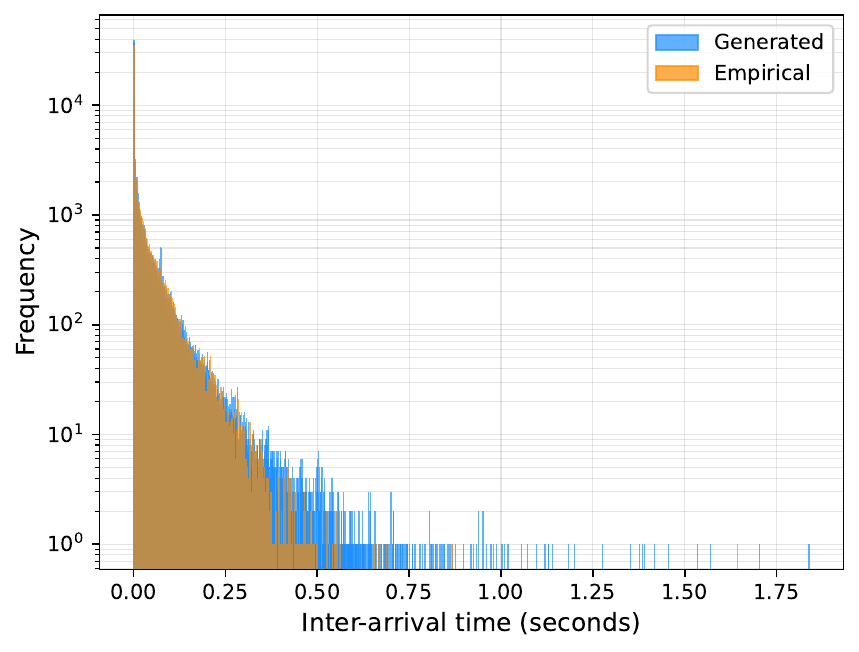}
         \caption{}
         \label{fig:cancel_inter}
     \end{subfigure}
     \hfill
     \begin{subfigure}{0.49\textwidth}
         \centering
         \includegraphics[width=\textwidth]{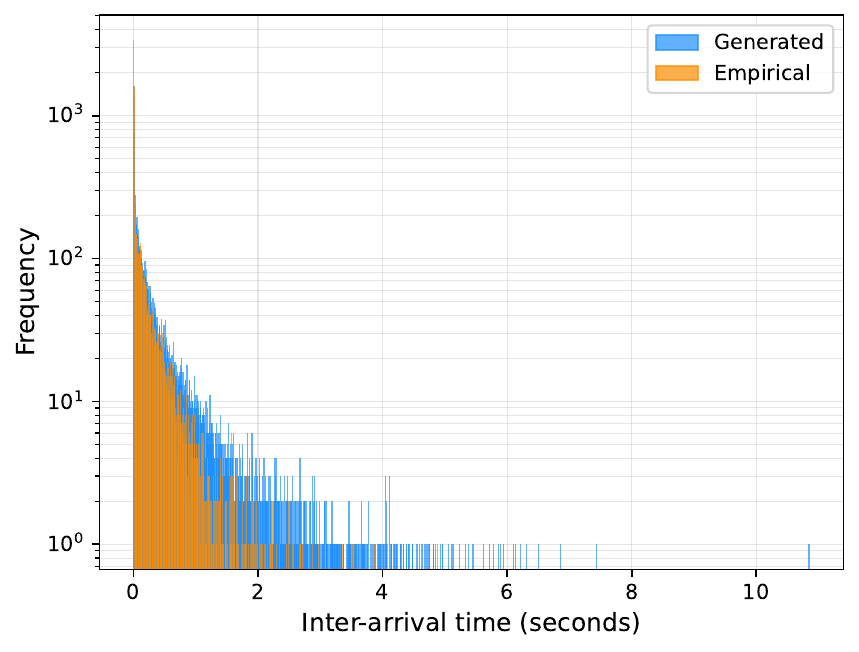}
         \caption{}
         \label{fig:replace_inter}
     \end{subfigure}
        \caption{The generated and empirical distributions of inter-arrival times on a semi-log scale (each histogram was composed of 500 bins). Data across all simulation trials (N=10) were similar and compared against the test distribution. Overlap between distributions (brownish color) is indicative of the model's ability to capture the placement rate of all order types. (a) The distribution of inter-arrival times for limit orders (add message). The data shows that all quotes are placed within one second of each other. (b) The distribution of inter-arrival times for market orders (execution message). Extreme outliers from a single trial were removed from the figure for perceptibility. (c) The distribution of inter-arrival times for cancel orders (deletion message). Both full cancellations and partial cancellations are included for simplicity. (d) The distribution of inter-arrival times for replace orders (replacement message).}
        \label{fig:inter_arrival}
\end{figure}

% order inter-arrival rates for each order type
The time elapsed between order placements was well approximated by the model for all message types. For all message types, our model performed better at capturing shorter time intervals rather than longer intervals, as evidenced by the slightly greater mass (blue bins) on the right tail of each message type distribution (Fig. \ref{fig:inter_arrival}). The model performed best at reproducing the inter-arrival times for add and cancel messages, which is likely due to the greater number of training examples for these message types. The model performed slightly worse at reproducing the inter-arrival times for execution and replace messages, which is likely due to the higher error rate of these message types. The model was able to capture the clustering of order placements at the sub-second level, which is a key feature of real order flow data.

\begin{figure}[ht]
    \centering
    \begin{subfigure}{0.49\textwidth}
        \centering
        \includegraphics[width=\textwidth]{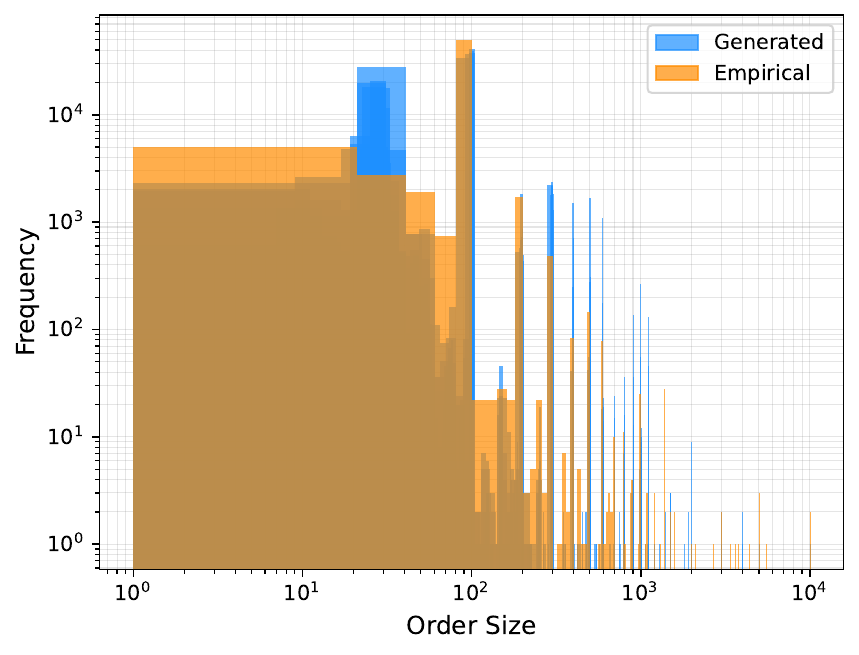}
        \caption{}
        \label{fig:add_size}
    \end{subfigure}
    \hfill
    \begin{subfigure}{0.49\textwidth}
        \centering
        \includegraphics[width=\textwidth]{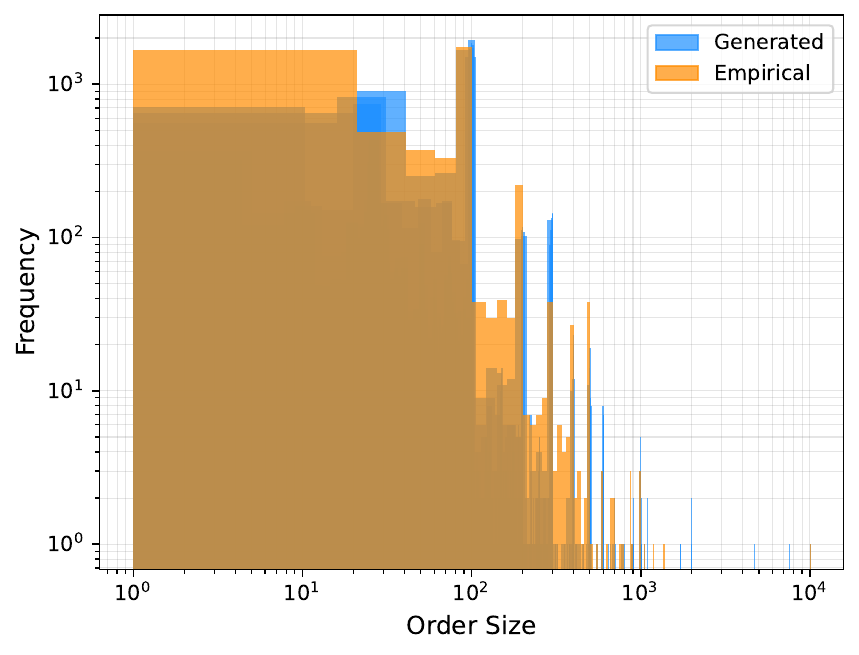}
        \caption{}
        \label{fig:exec_size}
    \end{subfigure}
    \hfill
    \begin{subfigure}{0.49\textwidth}
        \centering
        \includegraphics[width=\textwidth]{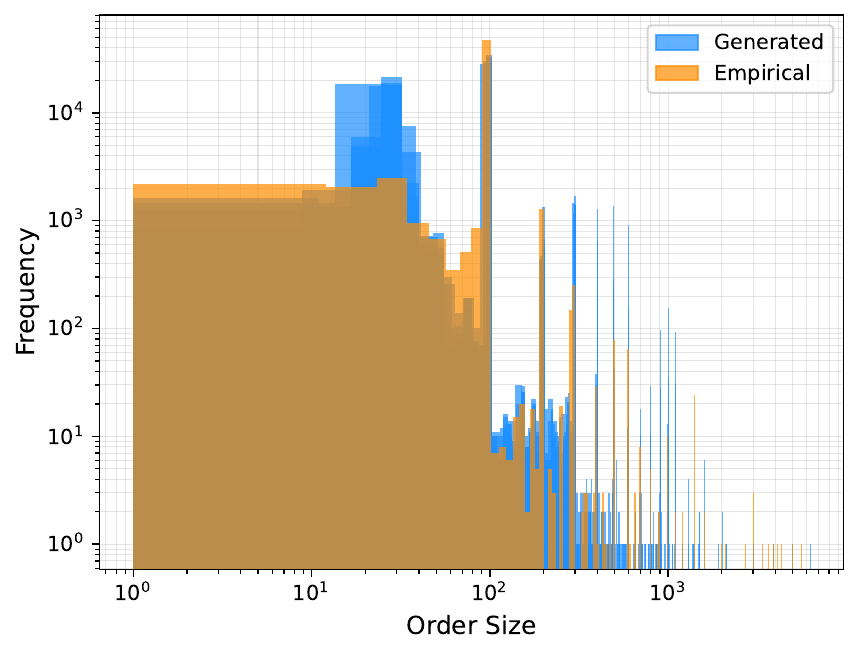}
        \caption{}
        \label{fig:cancel_size}
    \end{subfigure}
    \hfill
    \begin{subfigure}{0.49\textwidth}
        \centering
        \includegraphics[width=\textwidth]{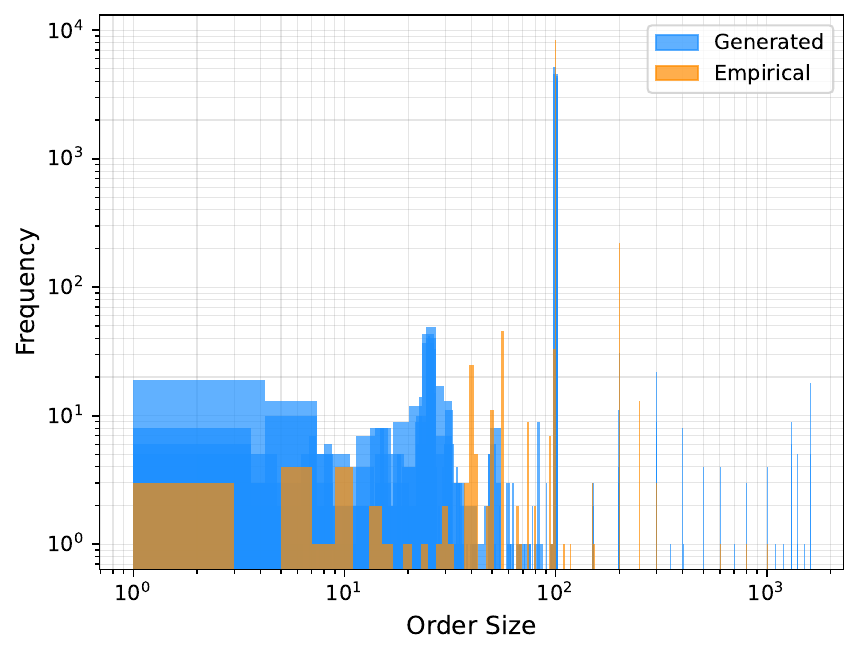}
        \caption{}
        \label{fig:replace_size}
    \end{subfigure}
       \caption{The generated and empirical distributions of order sizes on a log scale (each histogram was composed of 500 bins). Data across all simulation trials (N=100) were similar and compared against the test distribution. Overlap between distributions (brownish color) is indicative of the model's ability to capture the order fill size for all message types. (a) The distribution of fill sizes for limit orders (add message). The data shows the characteristic peak at round lot sizes. (b) The distribution of fill sizes for market orders (execution message). (c) The distribution of fill sizes for cancel orders (deletion message). (d) The distribution of fill sizes for replace orders (replacement message).}
       \label{fig:order_sizes}
\end{figure}

The model reproduced several key features of the order size distributions, but was inconsistent across message types. We found that the model performed best at reproducing the distribution of execution order sizes---although each generated distribution had similar traits (Fig. \ref{fig:order_sizes}). We found that each distribution was able to reproduce the clustering of round lot sizes (i.e., order sizes placed in multiples of 100) \cite{Challet2001}. Each generated distribution, to different extents, underestimated smaller limit order sizes ($<$100 shares), which were much more random in size and difficult to predict. Similarly, the model under-predicted order size values that were between round lot sizes. Since execution, cancel, and replace message types are all referential, it is likely that the under-estimation of certain limit order sizes had influenced the results of all other message types. The comparison of replace orders sizes were the worst among message types, and this is likely due to the reasons mentioned earlier regarding the high error rate of replacement messages. An interesting finding is that the results of some referential message types (e.g., cancel messages) were comparable to the other non-referential message types (e.g., add messages), even with the limited context length. % we discuss this later.... This result indicates that a long context length is not needed to....

\begin{figure}[ht]
    \centering
    \begin{subfigure}{0.60\textwidth}
        \centering
        \includegraphics[width=\textwidth]{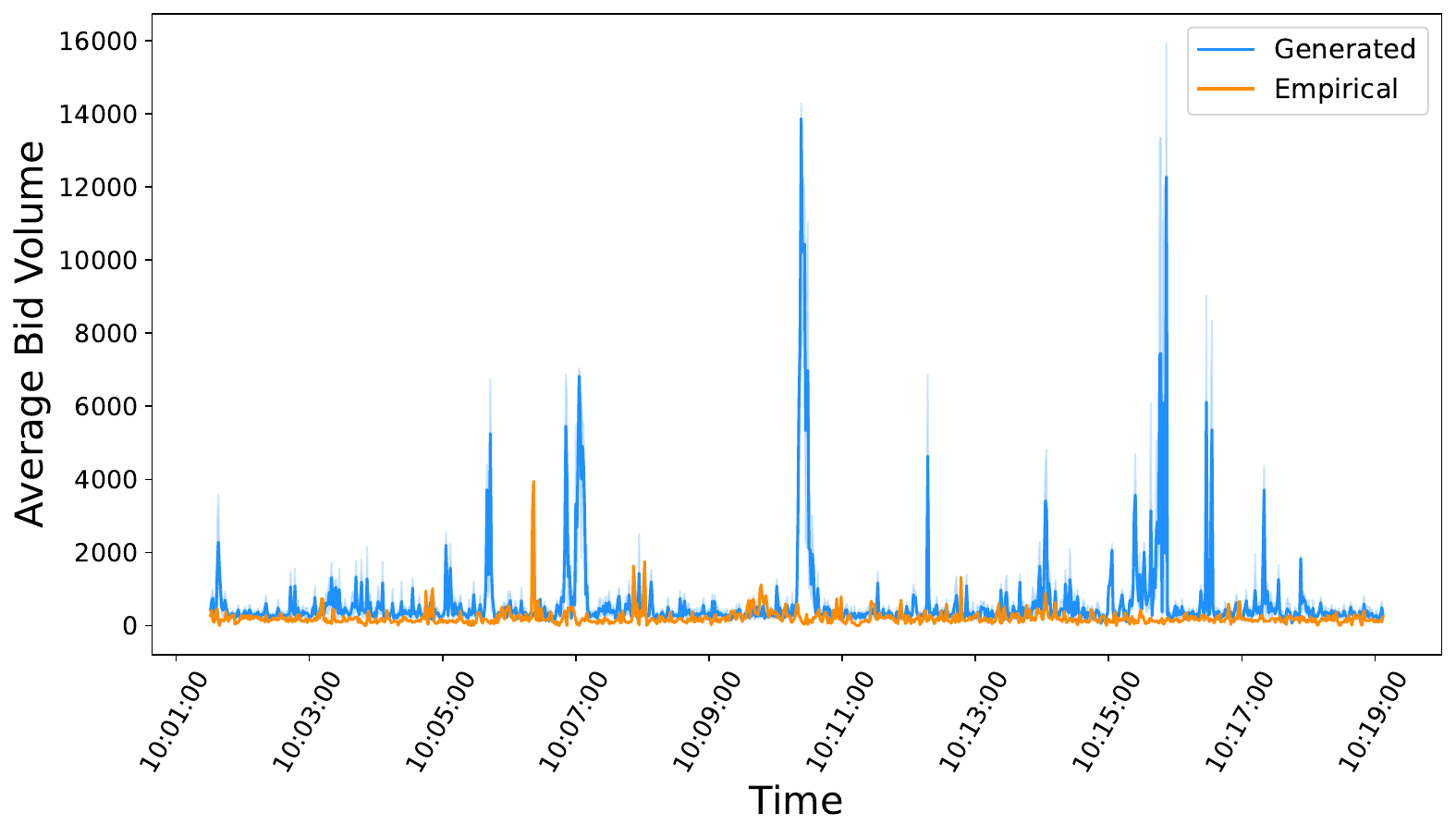}
        \caption{}
        \label{fig:bid_vol_1}
    \end{subfigure}
    \hfill
    \begin{subfigure}{0.60\textwidth}
        \centering
        \includegraphics[width=\textwidth]{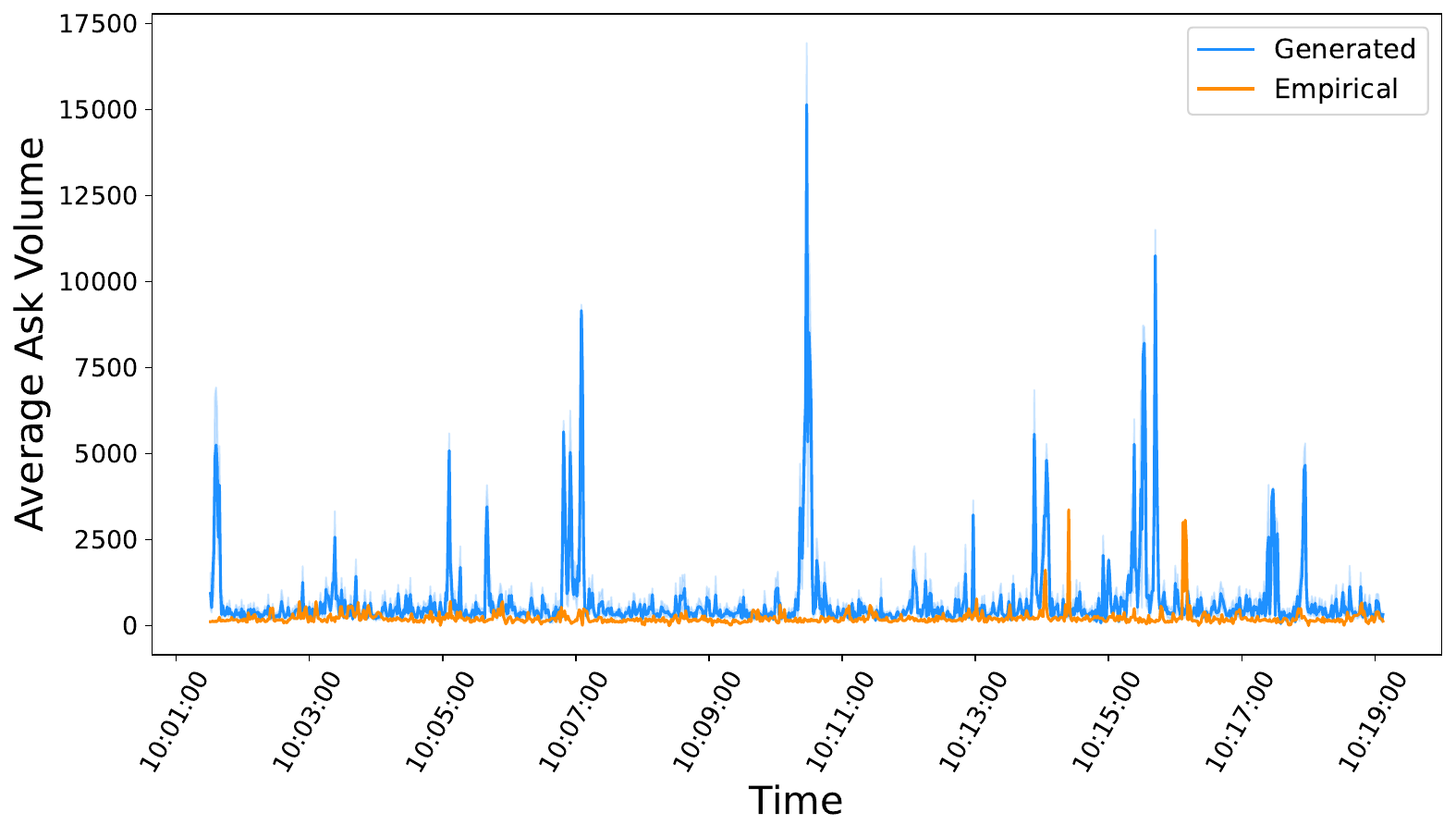}
        \caption{}
        \label{fig:ask_vol_1}
    \end{subfigure}
    \hfill
    \begin{subfigure}{0.60\textwidth}
        \centering
        \includegraphics[width=\textwidth]{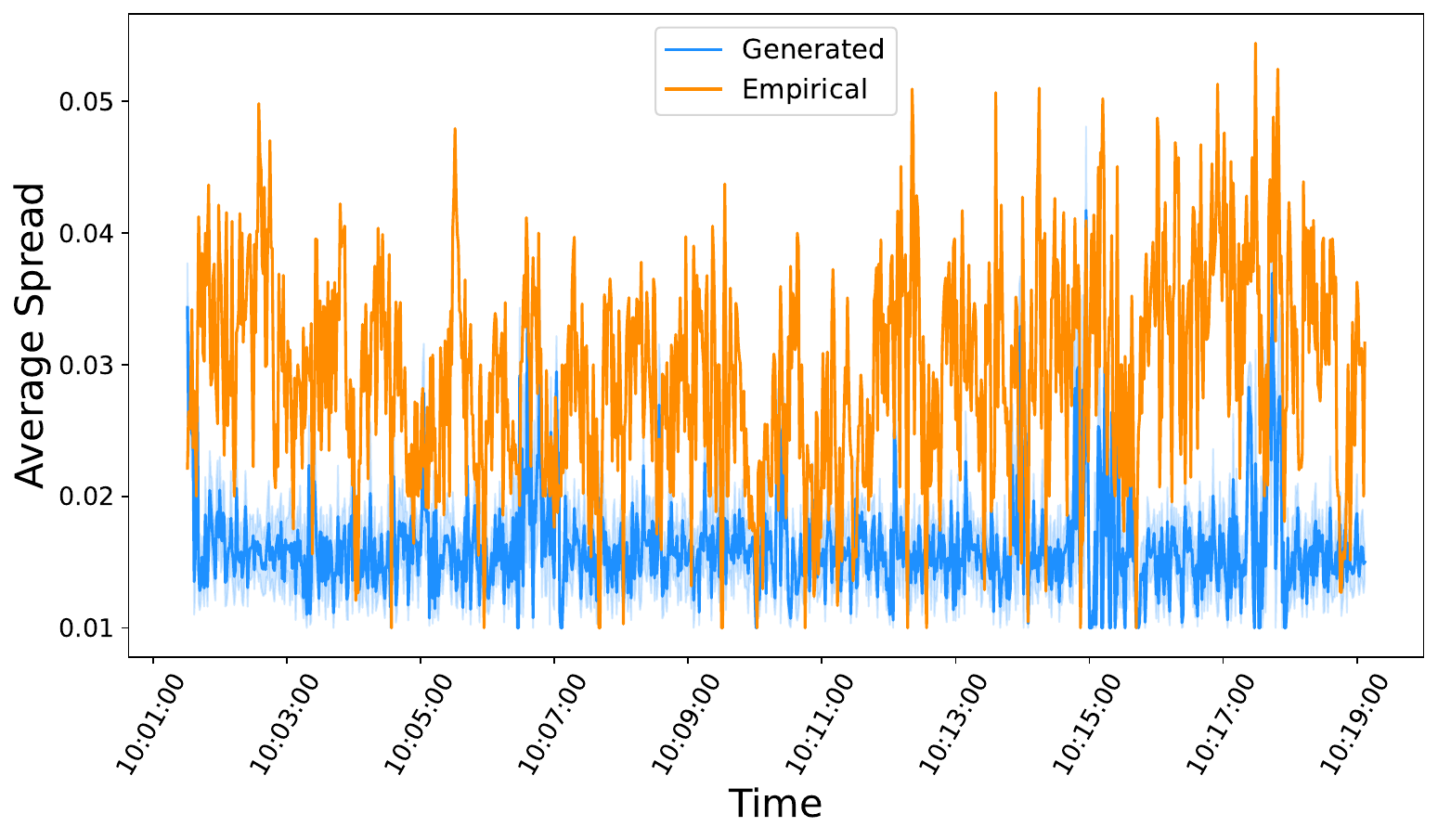}
        \caption{}
        \label{fig:gpt_spread}
    \end{subfigure}
       \caption{Spreads and volume offered at best bid and ask sides were averaged across 1-second intervals. (a) The average volume offered at the best bid price level. (b) The average volume offered at the best ask price level. (c) The spread, or difference between the best bid and ask price. Data across all simulation trials (N=10) were compared against the empirical test distribution (x-axis cutoff at the minimum final time value of generated sequences). The solid blue line denotes the ensemble mean of the generated average volume offered, the blue shaded region denotes the 95\% confidence region of the generated ensemble, and the solid orange line denotes the empirical average volume offered.}
       \label{fig:vol_bid_ask}
\end{figure}

The LOB simulator lacked realistic liquidity, which was evident in the average volume offered at the best bid $p^{b(1)}$ and ask price $p^{a(1)}$ levels (Figs. \ref{fig:bid_vol_1}, \ref{fig:ask_vol_1}) and the spread $s = p^{a(1)} - p^{b(1)}$ (Fig. \ref{fig:gpt_spread}). The model struggled to reproduce the average volume offered for both bid and ask sides of the LOB. The results showed periods where the generated average volume offered was only slightly higher than the test distribution, followed by periods where the offered volume rose to extreme values. This inconsistency was unsurprising given that the model does not have access to the book directly, rather just the previous context length messages. Related to this, the model under-predicted the average spread (albeit only by a few cents). This discrepancy in liquidity is arguably the most glaring issue with the LOB simulator and was the likely cause of other statistical deviations in the study.

\subsection*{Simulated properties of returns} % higher level stuff
% heavy tails
We investigated the generated price returns distribution for the existence of several well-known statistical regularities. Given a time scale $\Delta$, price returns are defined as
\begin{eqnarray}
    r_{t,\Delta} = \text{ln}(\frac{p_{t + \Delta}}{p_t})
\end{eqnarray}
where $p_t$ denotes the mid-price at time $t$, which is the mean of the best bid $p^{b(1)}_t$ and ask $p^{a(1)}_t$ price:
\begin{eqnarray}
    p_t = \frac{p^{b(1)}_t + p^{a(1)}_t}{2}
\end{eqnarray}

Our model reproduced the heavy-tailed property of the returns distribution, but with a noticeable difference in the peak at zero. The returns were bucketed into $\Delta=$ 1-second intervals to better visualize the high-peakedness and heavy tails of both empirical and generated distributions \cite{Mandelbrot1963} (Fig. \ref{fig:heavy_tails}). We quantified this property by measuring the kurtosis $\kappa$ of the returns distribution ($\kappa$=0 for a normal distribution). Excess kurtosis was present in both distributions, with $\kappa=2.44$ for empirical returns and a sample mean $\pm$ SD of $\kappa=4.93 \pm 1.99$ for generated returns. Although both distributions exhibited the heavy-tailed statistical property, they exhibited it to different degrees, with the generated returns having noticeably higher peaks at zero. This result is likely related to the model's imperfect fidelity of liquidity, which was evident in the average volume offered at the best bid and ask price levels.

% % heavy tails
% \begin{figure}[t]
%     % \centering
%     \includegraphics[width=0.99\textwidth]{Figures/marketGPT/heavy_tails_12302019_AAPL.pdf}
%     \caption{Distribution of 1-second returns of both generated and empirical messages. Data across all simulation trials (N=10) were compared against the test distribution. The solids lines denote the kernel density estimate for both generated (black line) and empirical (orange line) returns. Both distributions exhibited the heavy-tailed property--however, the generated returns distribution was noticeably more extreme.}
%     \label{fig:heavy_tails}
% \end{figure}

\begin{figure}[ht]
    \centering
    \begin{subfigure}{0.65\textwidth}
        \centering
        \includegraphics[width=\textwidth]{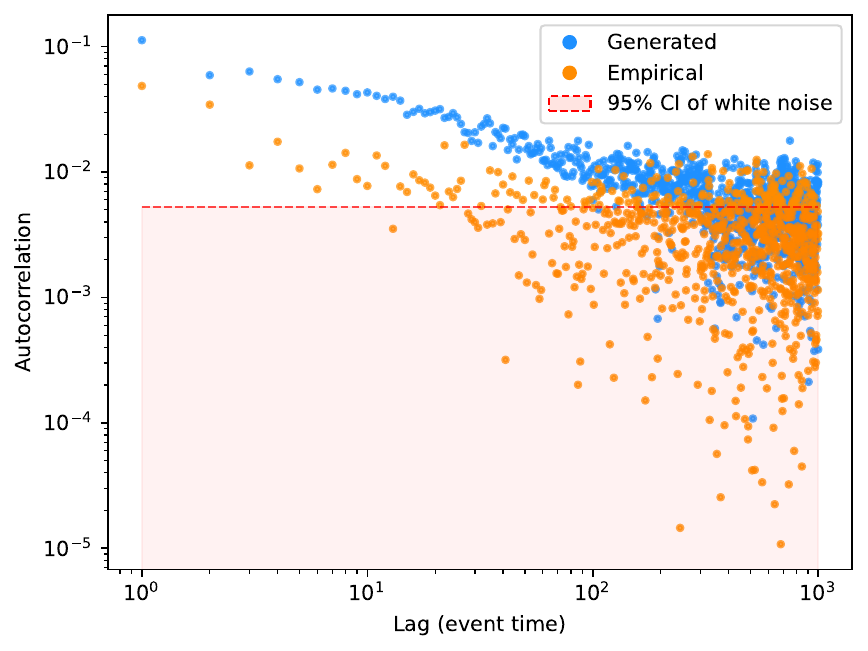}
        \caption{}
        \label{fig:autocorr_volatility}
    \end{subfigure}
    \hfill
    \begin{subfigure}{0.65\textwidth}
        \centering
        \includegraphics[width=\textwidth]{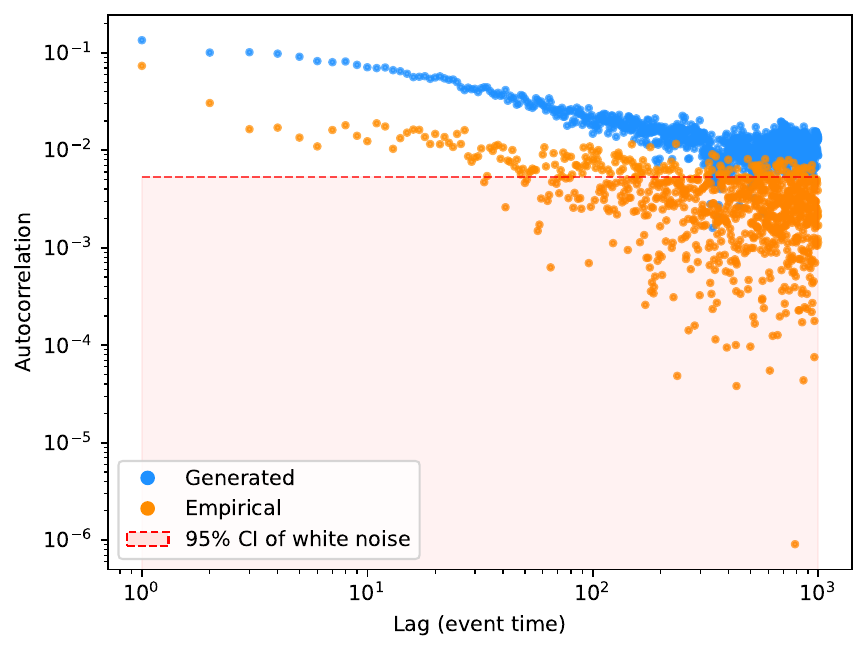}
        \caption{}
        \label{fig:autocorr_nonlin}
    \end{subfigure}
       \caption{Volatility clustering of generated and empirical returns. Data from a single trial is illustrated for clarity but the property was nearly identical across all simulation trials (N=10). (a) The autocorrelation of squared returns. (b) The autocorrelation of the absolute value of returns. Both returns series were plotted against a 95\% confidence region for white noise and exhibited significant positive autocorrelation for the initial lags before decaying.}
       \label{fig:vol_clustering}
\end{figure}

% volatility clustering and nonlinear dependence
Our model reproduced the complex and nonlinear nature of volatility clustering and long-range dependence in the returns series. We tested for the property of volatility clustering by measuring the autocorrelation of squared returns
\begin{eqnarray}
    C_{\text{sqr}} (\tau) = \text{corr} (r^{2}_{t+\tau, \Delta}, r^{2}_{t, \Delta})
\end{eqnarray}
where $\tau$ denotes the time lag \cite{Mandelbrot1963, Cont2001, Cont2005}. The results indicate that, like empirical volatility, our generated volatility series exhibited positive autocorrelation for several time lags and decayed slowly (Fig. \ref{fig:autocorr_volatility}). Similar results were observed for absolute returns (Fig. \ref{fig:autocorr_nonlin}). To further test for long-range dependence, we verified that the autocorrelation of absolute returns decayed according to a power law distribution:
\begin{eqnarray}
    C_{\text{abs}} (\tau) = \text{corr} (\left| r_{t+\tau, \Delta} \right|, \left| r_{t, \Delta} \right|) \simeq \tau^{-\gamma}
\end{eqnarray}
We used detrended fluctuation analysis (DFA) to measure the power law exponent of both empirical and generated absolute returns \cite{Peng1994, Hardstone2012}. The scaling exponent $\alpha$ is a generalization of the Hurst exponent and is $0.5 < \alpha < 1.0$ for time series that exhibit strong persistence, or momentum. We observed similar values for both series, which were estimated to be $\alpha = 0.64$ for empirical absolute returns and $\alpha = 0.73 \pm 0.02$ for generated absolute returns. The power law decay exponent $\gamma$ is related to $\alpha$ by $\gamma = 2 - 2(\alpha)$; accordingly, time series with $\gamma \in (0, 1)$ exhibit persistent long-range power-law correlations \cite{Buldyrev1995}. This corresponded to power law decay exponents $\gamma=0.72$ for empirical absolute returns and $\gamma = 0.54 \pm 0.05$ for generated absolute returns, which were similar to values reported in the literature \cite{Cont1997, Liu1997, Cont2001, Gopikrishnan1999}. We also estimated the Hurst exponent $H$ using Anis-Lloyd corrected rescaled range (R/S) analysis as an additional means of quantifying the volatility clustering and long-range dependence property along with 95\% confidence intervals \cite{Hurst1951, ANNIS1976, mandelbrot1971analysis}. An $H$ value well above 0.5 was estimated for both empirical $H=0.678$ and generated $H=0.77 \pm 0.05$ absolute returns, which was indicative of long-range dependence or \emph{persistent} behavior. Both values were observed to be well outside the 95\% confidence intervals for short-memory, e.g., white noise. Similar to the heavy tails property, for both volatility clustering and long-range dependence, we observed that our model exhibited the effect to a greater extent than what was observed empirically. This likely occurred for a number of reasons including the model's slight under-prediction of large execution order sizes. However, despite the slight deviation from empirical observation, the model demonstrated an impressive ability to reproduce all of these statistical regularities simultaneously. Interestingly, the model was able to exhibit slow decay of autocorrelation well beyond the context length of the model. This result was unexpected and we attributed it to the capacity of several transformer layers to learn highly abstract phenomena from the training data.

\subsection*{Predictive properties of the model}
\begin{figure}[ht]
    \centering
    \begin{subfigure}{0.49\textwidth}
        \centering
        \includegraphics[width=\textwidth]{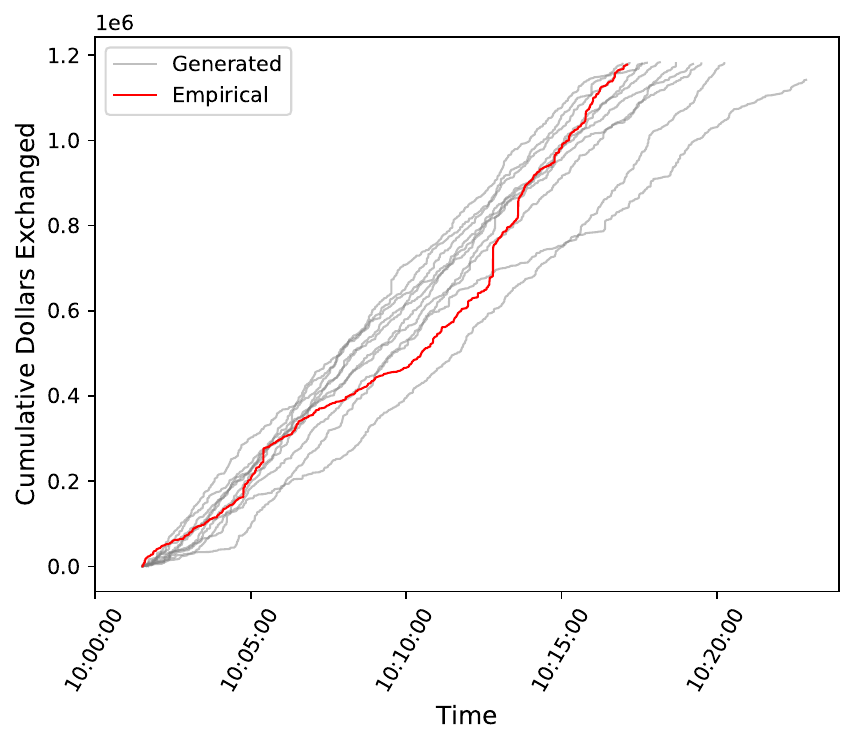}
        \caption{}
        \label{fig:cum_dollars}
    \end{subfigure}
    \hfill
    \begin{subfigure}{0.49\textwidth}
        \centering
        \includegraphics[width=\textwidth]{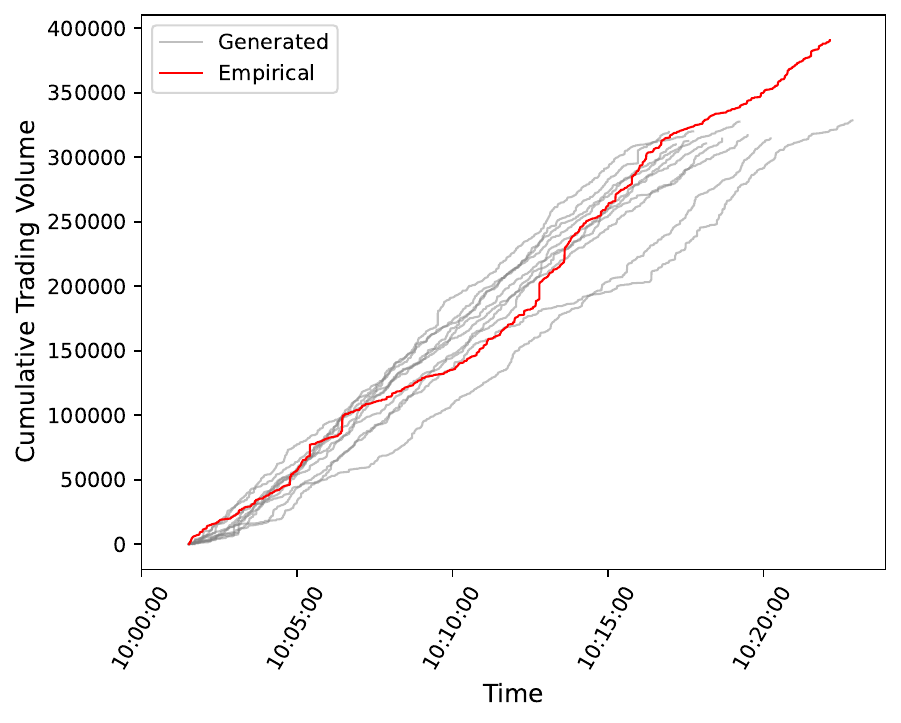}
        \caption{}
        \label{fig:cum_vol}
    \end{subfigure}
    \hfill
    \begin{subfigure}{0.49\textwidth}
        \centering
        \includegraphics[width=\textwidth]{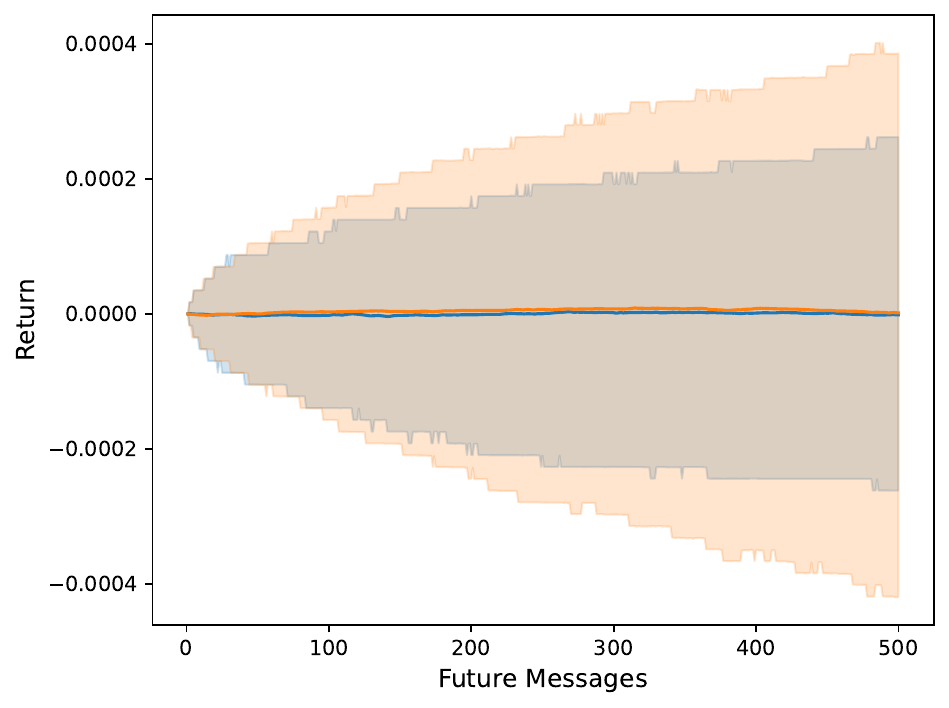}
        \caption{}
        \label{fig:future_returns}
    \end{subfigure}
    \hfill
    \begin{subfigure}{0.49\textwidth}
        \centering
        \includegraphics[width=\textwidth]{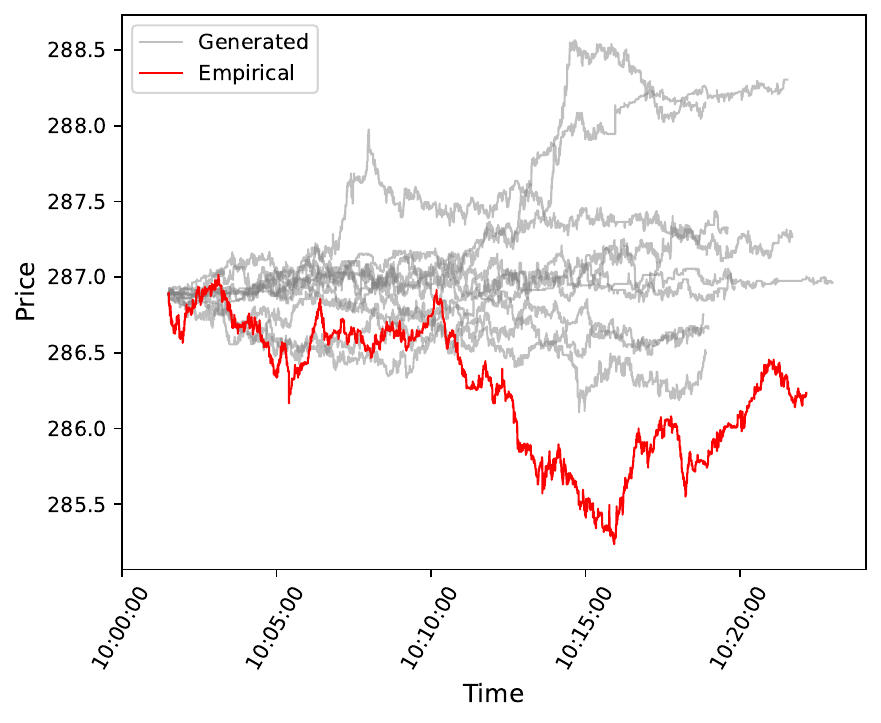}
        \caption{}
        \label{fig:price_traj}
    \end{subfigure}
       \caption{Prediction of (a) cumulative dollars and (b) shares traded for generated (grey) and empirical (red) time series data. The notional values of money ans shares exchanged were accurately predicted for most simulation trials. Each sequence is truncated by the minimum number of generated execution messages across simulation trials (4,116 messages). (c) Distribution of mid-price returns over the next 500 messages for empirical (blue) and generated (orange) data. Solid lines denotes the mean and the shaded regions cover 95\% of the distribution. Data from a single trial is illustrated for clarity but the property was nearly identical across all simulation trials (N=10). Returns were calculated between the mid-price $t$ future messages after, and the mid-price just before the start of the random sample. (n=1000 random samples drawn from both the empirical and generated distributions). (d) Price trajectories that emerged from generated messages and matched orders (grey) and the empirical price series (red). Data across all simulation trials (N=10) were compared against the empirical test distribution. Each sequence is truncated by the common sequence length (M=135,813 messages).}
       \label{fig:cum_traded}
\end{figure}

% returns vs future messages
The model roughly reproduced the distribution of returns computed over immediate and future messages. Returns were calculated between the mid-price just before the start of the random sample and the mid-price $t$ future messages after. We compared the distribution of mid-price returns over the next 500 messages for both empirical and generated data (Fig. \ref{fig:future_returns}). The results showed that the returns generated by our model, which overlapped approximately with the empirical distribution, did not show any signs of directional bias or trend across several sample sequences. This result was consistent across all simulation trials and was indicative of the model's ability to generate returns which were qualitatively similar to the empirical distribution over long time horizons.

% % returns vs future messages
% \begin{figure}[ht]
%     % \centering
%     \includegraphics[width=0.99\textwidth]{Figures/marketGPT/returns_vs_future_messages.pdf}
%     \caption{Distribution of mid-price returns over the next 500 messages for empirical (blue) and generated (orange) data. Solid lines denotes the mean and the shaded regions cover 95\% of the distribution. Data from a single trial (seed=42) is illustrated for clarity but the property was nearly identical across all simulation trials (N=10). Returns were calculated between the mid-price $t$ future messages after, and the mid-price just before the start of the random sample. (n=1000 random samples drawn from both the empirical and generated distributions)}
%     \label{fig:future_returns}
% \end{figure}

% forecasting capabilities of future messsages (still need to collect data for this) (save for next paper?)

% price & volume trajectory (and notional value diff)
The model was able to generate realistic price and volume trajectories. Since the model is trained solely on message data, rather than other pieces of information that drive price movement like news events, etc., we cannot claim that our model can predict future price or volume trajectories. However, we do claim that our model can produce highly realistic and plausible trajectories. To support this claim, in addition to the other results presented throughout this study, we include data that shows that our model closely approximates the notional value of money and shares exchanged over simulation time, i.e., the cumulative series of dollar value traded and trading volume (Figs. \ref{fig:cum_dollars}, \ref{fig:cum_vol}). We also include the price series of several trials that emerged as a result of the generated messages and order book matching (Fig. \ref{fig:price_traj}). These price series appear qualitatively similar (they share a distinct roughness and overall "Brownian" appearance) and stay within close region of one other. We found this result impressive considering that this model was not trained to forecast price series whatsoever, but rather predict the next token (message sub-element) in the sequence.

\subsection*{Model Limitations}
There are important considerations worth mentioning here that relate to this study's impact and practical use outside of the research setting. Firstly, the inference time of the model is quite high and limits downstream applications. As it stands, the model must generate 24 tokens per message, which is expensive considering that there are hundreds of thousands or even millions of messages in a single trading day. The autoregressive nature and quadratic complexity of the transformer architecture add to this difficulty and make it necessary to obtain advanced hardware to attempt to run large experiments with this type of model. The sampling parameters of the model are also difficult to tune and we had to rely on trial and error to find the best combination of temperature and top-p parameters. Further, the same parameters did not work well across models finetuned on other ticker symbols. This is a limitation of the model and is likely to be improved with the development of a more robust parameter selection procedure or more data and more parameters. 
% takes 4 hours to simulate ~20 mins of data on my machine

% fragmented market (other exchanges + dark pools + payment for order flow) and we only have Nasdaq (although this is the most frequently traded one)

% limited finetuning data
% limited compute so small model and limited context length
% poor inference times and inherent limitations due to autogressive nature of model, and how effects scaling to multilple assets?

The high computational cost of inference limited the number of samples and length of samples we could obtain in a reasonable amount of time. Ideally, we would like to extend this study to longer time sequences and to more finetuned ticker symbol models. We leave this to future work. Further, this study was conducted with consumer-grade hardware (a single NVIDIA RTX 4090), which limited the size of the model and the context length. Increasing the parameter and context size is likely necessary to explore multi-asset message generation and incorporating additional data sources as inputs to the model. Quantization methods may be explored as a means to reduce memory footprint and latency \cite{Gholami2021}. More advanced hardware is likely needed to scale the model, and alternative model architectures or extensions may need to be explored to mitigate the effects that this would have on the pre-training, finetuning, and inference cost. Speculative decoding has been proposed a means of accelerating inference of large Transformers, and low-rank model adaptations have been proposed as a storage- and compute-efficient alternative to full finetuning \cite{Leviathan2022, Hu2021, Dettmers2023}. We explored using an alternative tokenization procedure inspired by the Byte Pair Encoding (BPE) algorithm to reduce the number of tokens needed to represent a message, and although this successfully reduced inference times (by grouping frequent pairs of adjacent message sub-elements), we found that the model did not perform as well as with the tokenization scheme we used in this study. We believe this was due to the limited data (which was further minimized through BPE) and parameters used in this study, and that more data or more parameters may be needed to take full advantage of BPE tokenization.

% costly to generate sequences, which inhibits evaluation due to limited samples
From a research perspective, more evaluation can be done to survey the realism of the model. Market impact studies can be conducted to determine whether this model could be useful to optimal execution applications. Additionally, there are more statistical regularities to test for, such as the power law behavior of order lifetimes, intraday seasonality, and properties related to cross-asset correlations. These properties require longer sequences and improvements to be made to the model before they can be investigated, so they are left for future work. Recent work in recurrent architectures (and hybrid-recurrent architectures) have demonstrated strong performance on long-range tasks and may provide a promising solution for generating multi-asset messages over long periods \cite{Gu2023, Lieber2024, Ren2024}.
% save market impact for next paper
% mention tokenization here? mention how we tried it but it didn't perform very well---more data or more parameters may be needed to take full advantage of bpe
% mention difficulties of sampling parameters? Trial and error

\clearpage

\section*{Conclusions}
Building on recent work, we developed a simulation platform for generating order messages in a discrete event simulator. Our work leverages the power of deep learning to generate realistic order flow data that captures the complex and nonlinear nature of financial markets. This was possible by incorporating recent advancements in transformer-based language models, most notably the attention sinks which allowed us to condition the model on previous messages in a streaming fashion. Our model demonstrates the ability to produce realistic order flow data well after the initial prompt, which is a key feature for generating long sequences of plausible artifical market data. By implementing a model that produces order flow that is conditioned on previous messages, we allow future work on market impact studies where a user could interactively trade against the world agent in the DES. 

Our results show that our model generates highly realistic order flow, and to the best of our knowledge is the first study to show the validation of many of these stylized facts at the message-scale. We found that our model was able to reproduce several key features of the order flow data, including the distribution of message types, order sizes, and interarrival times. We also found that our model was able to reproduce the heavy-tailed property of the returns distribution, the volatility clustering of returns, and the power law behavior of absolute returns. We found that our model was able to generate realistic price and volume trajectories, and that the model was able to do all of this without these properties being explicitly part of the loss function. Our model was able to achieve all of this whilst being trained on modest hardware and on limited pretraining and finetuning data.

Additional work is needed to explore simultaneous multi-asset message generation and the inclusion of additional data sources. But given these promising results, further research and evaluation of these deep learning based bottom-up methods merit attention. Future work could investigate alternative model architectures and increased model and dataset sizes. We believe that this work is a step towards the development of more realistic and data-driven market simulators that can be used for a variety of applications in finance and beyond.

\clearpage

\section*{Conflict of Interest Statement}
The authors declare that the research was conducted without any commercial or financial relationships that could potentially create a conflict of interest.

\section*{Author Contributions}
J.V. directed the study, A.W. developed the model and codes, conducted the simulations and analysis, 
generated the figures. J.V. and A.W. edited the manuscript. All authors reviewed this manuscript.

\section*{Acknowledgments}
We gratefully acknowledge the suggestions from the anonymous reviewers to improve this manuscript. 

\section*{Data Availability Statement}
Model code is available under an MIT software license from the Varnerlab GitHub repository: \url{https://github.com/aaron-wheeler/MarketGPT}.

\clearpage

\bibliography{References_v1}

\clearpage

%%%%%%%%%%% FIGURES %%%%%%%%%%%%%

\clearpage

 % ...state variables are updated (e.g., sim time, book) and other orders are handled before cycle repeats

\clearpage

% order type percentages and heavy tails

\clearpage

% order inter-arrival rates

\clearpage

% order size distributions

\clearpage

% volume at best bid and ask and spread

\clearpage

 % include DFA fit? appendix?

\clearpage

% cumulative dollars and volume traded, returns vs future messages, price trajectory

\clearpage

% Supplemental figures -
% Set the S-
\renewcommand\thefigure{S\arabic{figure}}
\renewcommand\thetable{T\arabic{table}}
\renewcommand\thepage{S-\arabic{page}}
\renewcommand\theequation{S\arabic{equation}}
%
% Reset the counters -
\setcounter{equation}{0}
\setcounter{table}{0}
\setcounter{figure}{0}
\setcounter{page}{1}

% include the supplement -
% \include{Supplement}

\end{document}